\documentclass[preprint,3p,12pt]{elsarticle}
\usepackage{mathrsfs}
\usepackage{amsmath}
\usepackage{stmaryrd}
\usepackage{bbding}
\usepackage{dcolumn}
\usepackage{graphicx}	
\usepackage{amsfonts}
\usepackage{amssymb}
\usepackage{psfrag}
\usepackage{wrapfig}
\usepackage{subfigure}
\usepackage{makeidx}
\usepackage{bm}
\usepackage{epsf}
\usepackage{epsfig}
\usepackage{setspace}
\usepackage{graphicx}
\usepackage{epstopdf}
\usepackage{psfrag}
\usepackage{subfigure}
\usepackage{color}
\usepackage{comment}
\usepackage{caption}
\newtheorem{rmk}{Remark}

\renewcommand{\vec}[1]{{\boldsymbol{#1}}}

\begin{document}

\title{Wave–Particle Turbulent Simulation of Spatially Developing Round Jets Using a Non Equilibrium Transport Model with a Mixing Length Characteristic Time Closure}

\author[HKUST1]{Xiaojian Yang}
\ead{xyangbm@connect.ust.hk}

\author[HKUST1,HKUST2,HKUST3]{Kun Xu\corref{cor1}}
\ead{makxu@ust.hk}

\address[HKUST1]{Department of Mathematics, Hong Kong University of Science and Technology, Clear Water Bay, Kowloon, Hong Kong}
\address[HKUST2]{Department of Mechanical and Aerospace Engineering, Hong Kong University of Science and Technology, Clear Water Bay, Kowloon, Hong Kong}
\address[HKUST3]{Shenzhen Research Institute, Hong Kong University of Science and Technology, Shenzhen, China}
\cortext[cor1]{Corresponding author}

\begin{abstract}

In this paper, the wave–particle turbulent simulation (WPTS)—a recently developed multiscale, non-equilibrium turbulence modeling approach—is coupled with a turbulence characteristic-time closure derived from Prandtl’s mixing-length hypothesis and applied to spatially developing round jets. In WPTS, fluid elements in strongly turbulent regions are represented by Lagrangian particles that travel a finite distance before interacting with the background flow field represented in a wave-like (Eulerian) form. This mechanism bears conceptual similarity to the discrete fluid parcels invoked in the Prandtl mixing-length picture.
WPTS differs from conventional mixing-length-based turbulence models in two key respects. First, particle evolution follows a non-equilibrium transport mechanism, rather than the equilibrium assumptions typically embedded in eddy-viscosity closures. Second, WPTS advances the wave and particle components in a coupled manner, with the particle fraction governed primarily by the modeled turbulence characteristic time, enabling laminar and turbulent regimes to be represented within a unified framework. Because spatially developing jets provide a canonical test case with well-established similarity behavior, they are used here for evaluation. Specifically, this work (1) develops a mixing-length-based characteristic-time model tailored to jet flows and (2) incorporates it into WPTS to assess predictive performance. The resulting WPTS framework accurately reproduces the jet similarity solution and other characteristic features at Reynolds numbers of 5,000 and 20,000, demonstrating the promise of WPTS as a practical tool for turbulence modeling and simulation.

\end{abstract}

\begin{keyword}
wave-particle turbulent simulation, non-equilibrium transport, jet flow
\end{keyword}

\maketitle

\section{Introduction}	

Turbulence modeling and simulation are essential not only for advancing fundamental understanding of turbulent dynamics but also for reliably predicting flow fields in practical engineering applications \cite{pope2001turbulent}. Over the past decades, extensive research has established several widely used approaches, including direct numerical simulation (DNS), which solves the Navier–Stokes (NS) equations without turbulence modeling; large eddy simulation (LES), which resolves the large-scale motions while requiring a closure for the subgrid-scale stresses; and Reynolds-averaged Navier–Stokes (RANS) methods, in which additional modeled relations and/or transport equations are introduced to close the Reynolds-stress terms.
Among the earliest and most influential ideas in turbulence modeling is Prandtl’s mixing-length hypothesis (1925) \cite{Tur-model-mixlen-ori-prandtl19257}. It postulates that, in a turbulent flow, discrete fluid eddies—analogous to gas molecules—travel a finite distance while retaining their momentum characteristics before exchanging momentum with the surrounding fluid. The statistically averaged travel distance is termed the mixing length, denoted $l^*$.
For simple shear flows, the turbulent (eddy) viscosity can then be estimated as the product of a characteristic fluctuating velocity scale $v^*$
and the mixing length $l^*$.
Subsequent developments embedded this concept within the Boussinesq approximation, leading to the classical mixing-length turbulence model—a representative zero-equation eddy-viscosity closure widely used in RANS computations.
In addition to the original formulation, numerous variants have been proposed to improve predictive accuracy and extend applicability across a broader range of flows \cite{Tur-model-mixlen-wall-pirozzoli2014revisiting}. A notable example is the Cebeci–Smith (CS) model \cite{Tur-model-mixlen-wall-cebeci2012analysis}, which accounts for inner/outer-layer structure and has demonstrated good performance for wall-bounded turbulent flows. More broadly, mixing-length concepts and their extensions have been adopted in diverse areas, including oceanography \cite{Tur-model-mixlen-shallow-stansby2003mixing}, atmospheric science \cite{Tur-model-mixlen-atmospheric-huang2013turbulence, Tur-model-mixlen-storms-hanley2015mixing}, and other fields \cite{Tur-model-mixlen-stellar-joyce2023review}.

Unlike conventional turbulence models formulated within an eddy-viscosity framework, the kinetic-based wave–particle turbulent simulation (WPTS) method has recently been proposed and has shown promising performance for several canonical turbulent-flow problems \cite{Tur-wpts-first-yang2025wave, Tur-wpts-second-yang2025wave}. More broadly, turbulence modeling based on kinetic theory has been actively studied and exhibits substantial potential \cite{Tur-kinetic-chen2003extended, Tur-kinetic-luan2025constructing, Tur-kinetic-xin2025model}. A key distinction of kinetic-theory-based approaches is that they operate on the distribution function rather than directly on macroscopic variables, thereby providing additional modeling flexibility and a natural multiscale description.
WPTS is developed from the unified gas-kinetic scheme (UGKS) \cite{UGKS-xu2010unified, UGKS-book-xu2014direct} and its unified gas-kinetic wave–particle (UGKWP) variant. These methods were originally designed for rarefied-gas dynamics \cite{WP-first-liu2020unified, WP-second-zhu-unstructured-mesh-zhu2019unified} and have since been extended to a wide range of multiscale transport problems, including radiation transport \cite{WP-radiation-implicit-liu2023implicit, WP-three-photon-transport-li2020unified, WP-radiation-mf-yang2025unified}, plasmas \cite{WP-four-liu2020unified, WP-plasma-pip-pu2025unified}, and multiphase flows \cite{WP-six-gas-particle-yang2021unified, WP-gasparticle-polydis-yang2024unified}, among others.
WPTS is specifically intended for turbulence simulations on coarse grids, with resolutions significantly lower than those required for DNS. Its central feature is the coupled evolution of wave and particle components. The wave (Eulerian) component represents the large-scale flow structures resolved by the coarse mesh, whereas stochastic Lagrangian particles describe the dynamics of unresolved small-scale motions within a computational cell, such as the unresolved turbulent kinetic energy (TKE). In regions of strong turbulence, particles are sampled and advected downstream with characteristic velocities; their transport distance is controlled by a local turbulent characteristic time scale provided by the $\tau_t$ model. After transport, the particles are removed and their conserved quantities are deposited back into the background flow represented by the wave component, thereby completing the wave–particle coupling.

Consequently, the particle treatment in WPTS bears conceptual resemblance to the motion of discrete fluid parcels envisioned in Prandtl’s mixing-length hypothesis and related models. Nevertheless, WPTS differs fundamentally from traditional mixing-length turbulence closures in two key respects. First, WPTS explicitly resolves the non-equilibrium transport of particle-represented fluid elements, whereas mixing-length models act primarily through local modifications of the turbulent (eddy) viscosity in the diffusive terms. This enables WPTS to naturally incorporate non-local effects, because particles can travel across multiple cells and transport flow information over finite distances \cite{Tur-model-comment-lumley1992some}. Second, particle evolution in WPTS is dynamically coupled with the background wave component, providing a unified wave–particle description across laminar-to-turbulent regimes.
Compared with turbulence approaches based solely on Lagrangian particles, WPTS further improves efficiency by automatically reducing the particle fraction in regions where turbulence weakens. In the laminar limit, the method reverts to the underlying gas-kinetic scheme (GKS) and recovers the Navier–Stokes solution \cite{GKS-2001}.

Spatially developing round jets constitute a canonical turbulence problem and have been extensively investigated. A jet convects downstream while spreading radially: a high-velocity core gradually weakens with distance from the axis, transitioning toward a relatively quiescent outer flow \cite{pope2001turbulent}. Moreover, extensive experimental and numerical evidence indicates that fully developed turbulent jets exhibit robust similarity behavior over a range of Reynolds numbers, including the decay of the centerline velocity and self-similar profiles of key statistical quantities at different downstream stations \cite{Tur-case-jet-exp-hussein1994velocity, Tur-case-jet-exp-panchapakesan1993turbulence, Tur-case-jet-DNS-sharan2021investigation}. These well-established features make the round jet a natural and stringent benchmark for evaluating new turbulence modeling approaches.
Accordingly, this study develops a mixing-length-based model for the turbulence characteristic time scale, $\tau_{t}$, tailored to spatially developing jets, and integrates it into WPTS to assess the capability of WPTS to predict jet turbulence accurately.

In our previous work, WPTS was applied to a round jet at Re=5000 using a turbulence characteristic-time model $\tau_{t}$ that relies primarily on the Smagorinsky model (SM) \cite{Tur-wpts-second-yang2025wave}, which is widely used in LES. In the present study, we further examine WPTS coupled with a $\tau_{t}$ model derived directly from Prandtl’s mixing-length hypothesis. This choice is motivated by the prospect of constructing a characteristic-time closure with weaker Reynolds-number dependence, thereby improving the robustness of the WPTS framework across different flow conditions. As an initial step, we focus on accurately reproducing the key similarity behavior observed in the fully developed turbulent region of round jets.
We consider jets at Reynolds numbers $5,000$ and $20,000$. Although Re=$20,000$ is still moderate relative to many practical applications, it remains challenging for high-fidelity numerical simulation. Due to the prohibitive computational cost, DNS of spatially developing jets is generally restricted to relatively low Reynolds numbers. For instance, Shin et al. reported DNS up to Re=7290 \cite{Tur-case-jet-DNS-7290-shin2017self}, which is among the highest Reynolds numbers achieved in DNS studies of this problem. LES can reach higher Reynolds numbers at lower cost; however, there is no universally accepted grid-resolution standard for jet LES. A survey of the literature suggests that, for jets at Re $\simeq 10,000$, LES typically employs on the order of $10^7 $ cells \cite{Tur-case-jet-les-bogey2009turbulence, Tur-case-jet-les-rediff-bonelli2021high, Tur-case-jet-les-kim2009large}, still implying substantial computational expense. Against this background, Re=$20,000$ provides a meaningful and practical target for assessing the performance of WPTS in this study under a mesh points of RANS level.

The remainder of this paper is organized as follows. Section 2 describes the WPTS numerical algorithm and the mixing-length-based turbulence characteristic-time model, $\tau_{t}$. Section 3 presents WPTS predictions and accompanying analysis for spatially developing jet simulations performed on coarse grids of approximately $0.5 \times 10^6$ cells for Re=5000 and $1.0\times 10^6$ cells for Re=20000. Finally, Section 4 summarizes the main conclusions of the study.

\section{Wave-Particle Turbulent Simulation (WPTS) Method}
Constructed within a finite-volume method (FVM), WPTS constitutes a coupled wave-particle evolution framework, specifically designed for turbulence modeling and simulation under coarse grids, i.e., whose resolution is insufficient for DNS requirements.
As the name implies, WPTS comprises two components: the wave component, which represents the grid-resolved flow structures, while the stochastic particle models the evolution features of unresolved flow structures under the employed coarse mesh, typically the unresolved turbulent kinetic energy (TKE).
Specifically, the governing equation is based on the kinetic equation with a BGK relaxation model, i.e.,
\begin{gather}\label{bgk}
\frac{\partial f}{\partial t}
+ \nabla_x \cdot \left(\vec{u}f\right)
= \frac{g-f}{\tau},
\end{gather}
where $f$ is the probability distribution function (PDF) of fluid elements, $\vec{u}$ is the velocity, $\tau$ is the characteristic time, and $g$ is the equilibrium state with
\begin{gather}\label{geq}
g = \rho \left(\frac{\lambda}{\pi} \right)^{\frac{K+3}{2}} e^{-\lambda \left[\left(\vec{u} - \vec{U}\right)^2 + \vec{\xi}^2 \right]}.
\end{gather}
$K$ is the internal degree of freedom taken as 2 for the diatomic molecule gas.

In WPTS, the weighting between wave and particle components within a grid cell is mainly determined by the local turbulence intensity of the flow field. In regions where particles vanish, the fluid naturally remains in a laminar state.
Consequently, WPTS achieves the unified description spanning laminar to turbulent flow.
With this modeling concept in mind, we know $\lambda = \frac{1}{2RT}$ where $T$ stands for the thermal temperature $T_{thermal}$ only for laminar, but it also includes the turbulence temperature $\Theta_t$ in the case of turbulence,
where $\Theta_t$ is uniquely determined by the TKE, $\rho E_t$, through the relation $\rho E_t \overset{def}{=} \frac{3}{2}\rho \Theta_t$.
Accordingly, the total energy $\rho E$ for laminar flow includes the kinetic energy $\frac{1}{2}\rho \vec{U}^2$ and thermal energy $\rho R T_{thermal}$, while for turbulence it also includes $\rho E_t$ as well.
It is worth noting that, take moments with $g$ by $\vec{\psi}$ and then the macroscopic conservative variables, $\vec{W}=\left(\rho, \rho \vec{U}, \rho E\right)^T$, can be obtained, where $\vec{\psi}=(1,\vec{u},\displaystyle \frac{1}{2}\left(\vec{u}^2+\vec{\xi}^2\right))^T$.
The conservation law is satisfied based on the compatibility condition

\begin{gather*}
\int \vec{\psi}\left(g - f\right)\text{d}\vec{\Xi} = \vec{0},
\end{gather*}
where $\text{d}\vec{\Xi}=\text{d}u\text{d}v\text{d}w\text{d}\xi_1...\text{d}\xi_{K}$.

Next, the construction of WPTS will be introduced in detail.
Starting from the kinetic equation with the BGK relaxation term, the integral solution can be obtained,
\begin{equation}\label{bgk-integrasol}
f(\vec{x},t,\vec{u})=\frac{1}{\tau}\int_0^t g(\vec{x}',t',\vec{u} )e^{-(t-t')/\tau}\text{d}t'\\
+e^{-t/\tau}f_0(\vec{x}-\vec{u}t, \vec{u}),
\end{equation}
where $\vec{x}'=\vec{x}+\vec{u}(t'-t)$ is the trajectory of particles, $f_0$ is the initial gas distribution function at time $t=0$.
The integral solution above illustrates that the evolution process comprises two components: the cumulative effect of the equilibrium state and the free transport of the initial distribution function.
This characteristic aligns reasonably with the feature of the so-called non-local turbulence modeling, where the upstream flow information undergoes attenuation during transmission, and consequently, the effect on the local flow is a weighted combination of the flow information from different upstream locations.

In the WPTS framework, the contribution from the equilibrium state is represented by the hydrodynamic wave resolved by the employed grid. Specifically, we employ the NS equations to represent this part, ensuring that the WPTS formulation fully reverts to the NS solution in regions devoid of particles.
Further detailed investigations into the modeling choices for this component and their impact on the final results remain a subject for future research and fall beyond the scope of this paper.
The gas-kinetic scheme (GKS) is a kinetic solver targeting the NS equation. Consequently, the $\vec{F}^{eq}$ in WPTS employs the formulation in GKS, namely, the distribution function $f$ at cell interfaces is constructed based on Taylor expansion in spatial and temporal space, from which the equilibrium flux can be derived.
Generally, the calculation of this part can be summarized as three steps, as the standard procedure in FVM:
\begin{enumerate}
	\item Reconstruction: interpolating cell-averaged conserved variables $\vec{W}$ to interfaces. In this paper, the fifth-order WENO-AO reconstruction is employed.
	\item Evolution: computing fluxes $\vec{F}^{eq}$ by taking moment of $f^{eq}$,
	\begin{gather}\label{eqFluxeq}
	\vec{F}^{eq}_{ij}
	=\frac{1}{\Delta t} \int_{0}^{\Delta t} \int \vec{u}\cdot\vec{n}_{ij} f_{ij}^{eq}(\vec{x},t,\vec{u})\vec{\psi}\text{d}\vec{u}\text{d}t,
	\end{gather}
	where $f^{eq}$ is the time-dependent PDF of the equilibrium state at the interface,
	\begin{gather}\label{eqfeq}
	f^{eq}(\vec{x},t,\vec{u}) \overset{def}{=} \frac{1}{\tau}\int_0^t g(\vec{x}',t',\vec{u})e^{-(t-t')/\tau}\text{d}t' \nonumber\\
	= c_1 g_0\left(\vec{x},\vec{u}\right)
	+ c_2 \overline{\vec{a}} \cdot \vec{u} g_0\left(\vec{x},\vec{u}\right)
	+ c_3 A g_0\left(\vec{x},\vec{u}\right),
	\end{gather}
	with coefficients,
	\begin{align}\label{coefeq}
	c_1 &= 1-e^{-t/\tau}, \notag \\
	c_2 &= \left(t+\tau\right)e^{-t/\tau}-\tau, \\
	c_3 &= t-\tau+\tau e^{-t/\tau} \notag .
	\end{align}	
	\item Projection: updating the cell-averaged conserved variables $\vec{W}$ by $\vec{F}^{eq}$. Since both the equilibrium and free-transport fluxes are required in the projection, the complete formula for updating the conserved variables ($\vec{W}$) will be given later in the detailed presentation of all flux contributions.
\end{enumerate}

The second component $e^{-t/\tau}f_0(\vec{x}-\vec{u}t, \vec{u})$ in Eq.(\ref{bgk-integrasol}) constitutes the free transport mechanism. Based on our experience of the problems involved in non-equilibrium transport, this component is critical. Within the WPTS framework, this component is simulated through the wave-particle formulation. Particularly, the stochastic particles represent the discrete fluid elements under the breakdown assumption, and therefore the stochastic Lagrangian particles are employed to model the turbulence flow structures within the unresolved coarse grid \cite{Tur-wpts-first-yang2025wave}.

To elucidate clearly the role of particles within the WPTS, a concise description is provided as follows. In WPTS, the particle inherently serves a dual role.
Firstly, the particles constitute a physical portion of the fluids, and together with the wave component give a complete description of the flow evolution.
Secondly, since the particle owns its characteristic velocity, which may differ from that of the cell-averaged value, it inherently carries information related to the evolution of unresolved TKE (under the coarse grid).

Specifically, the governing equations for these particles adopt a relaxation-type formulation with the external driving force, namely
\begin{gather}\label{dudtpar}
\frac{\text{d} \vec{u}\left(t\right)}{\text{d} t} = \frac{\vec{U} - \vec{u}}{\tau_n} + \vec{a}.
\end{gather}
In this paper, $\vec{a}$ is taken as $-\frac{\nabla p}{\rho}$, considering it as the dominant term, although $\vec{a}$ should in principle incorporate the effects of both the pressure gradient and the viscous terms from the NS equations.
The collision/characteristic time $\tau_n$ in Eq.\eqref{dudtpar} is taken as the sum of physical and turbulence values, namely, $\tau_n = \tau + \tau_t$, where $\tau = \mu/p$ is the physical one, and $\tau_t$ stands for the modeled turbulence characteristic time, which is constructed from the mixing-length hypothesis in this paper and will be introduced later.
Leveraging the UGKWP method, only the particles those are capable of moving statistically in one whole time step are sampled and evolved via Lagrangian particles, and the flux contribution of remaining particles whose transport time are smaller than one time step will be calculated by the wave formulation, denoted as $\vec{F}^{fr,wave}$ \cite{WP-first-liu2020unified, Tur-wpts-first-yang2025wave, UGKS-book-framework-xu2021cambridge}
\begin{align*}\label{eqFluxfrwave}
\vec{F}^{fr,wave}_{ij}
&=\vec{F}^{fr,UGKS}_{ij}\left(\vec{W}_i^{h}\right) - \vec{F}^{fr,DVM}_{ij}\left(\vec{W}_i^{hp}\right)\\
&=\frac{1}{\Delta t}\int_{0}^{\Delta t} \int \vec{u} \cdot \vec{n}_{ij} \left[ e^{-t/\tau}f_0(\vec{x}-\vec{u}t,\vec{u})\right] \vec{\psi} \text{d}\vec{u}\text{d}t\\
&-e^{-\Delta t/\tau_n} \frac{1}{\Delta t} \int_{0}^{\Delta t} \int \vec{u} \cdot \vec{n}_{ij} \left[g_0^h\left(\vec{x},\vec{u} \right) - t\vec{u} \cdot g_\vec{x}^h\left(\vec{x},\vec{u} \right) \right] \vec{\psi}\text{d}\vec{u}\text{d}t\\
&=\frac{1}{\Delta t} \int \vec{u} \cdot \vec{n}_{ij} \left[ \left(q_4  - \Delta t e^{-\Delta t/\tau_n}\right) g_0^h \left(\vec{x},\vec{u} \right)
+ \left(q_5 + \frac{\Delta t^2}{2}e^{-\Delta t/\tau_n}\right) \vec{u} \cdot g_\vec{x}^h\left(\vec{x},\vec{u} \right) \right]\vec{\psi}\text{d}\vec{u},
\end{align*}
with
\begin{gather*}
q_4=\tau\left(1-e^{-\Delta t/\tau}\right), ~
q_5=\tau\Delta te^{-\Delta t/\tau} - \tau^2\left(1-e^{-\Delta t/\tau}\right).
\end{gather*}
The calculation of $\vec{F}^{fr,wave}$ follows the same procedure with $\vec{F}^{eq}$ described above, namely the reconstruction, evolution and projection. For the $\vec{W}^{h}$, the second-order reconstruction based on van Leer limiter is employed.

The evolution of stochastic particles is dominant in capturing the non-equilibrium transport in WPTS, and generally, it can be summarized as the following three steps:
\begin{enumerate}
	\item Sample particles from the wave,
	\begin{gather}\label{whp}
	\vec{W}^{hp}_i = e^{-\Delta t/\tau_n} \vec{W}^{h}_i,
	\end{gather}
	with $t_f = \Delta t$, and the velocity of the sampled particle is determined by
	\begin{gather}
	\vec{u}_p = \delta \vec{u}_p + \vec{U}
	\end{gather}
	where $\delta\vec{u}_p = \mathcal{D}_{N} \left[\rho E_t^{prod}, \rho^h\right]$, depending on the fluid density by wave component $\rho^h$, and the modeled production of TKE, which is assumed as $\rho E_t^{prod} = C_0\left(1-e^{-\Delta t/\tau_n}\right)\rho E_t$. In fact, since only the fluid corresponding to $\rho^{hp}$ in $\rho^{h}$ is sampled and represented in particle form as given in Eq.\eqref{whp}, overall the TKE production $e^{-\Delta t/\tau_n} \rho E_t^{prod}$ is loaded onto the newly sampled particles. The remaining value is carried by the wave component and will fully decay over one subsequent time step.
	More details about the particle sampling can be found in \cite{Tur-wpts-first-yang2025wave}.
	
	Determine the $t_f$ for the surviving particles from the last step (if existing),
	\begin{gather}
	t_f = \text{min}\left[-\tau_n\text{ln}\left(\eta\right), \Delta t\right].
	\end{gather}
	\item Move particles by operator splitting, which indicates,
	\begin{gather}
	\vec{x}^* = \vec{x}^n + \vec{u}^n t_f,
	\end{gather}
	for free streaming, and
	\begin{align}
	\vec{u}^{n+1} &= \vec{u}^n + \vec{a} t_f, \\
	\vec{x}^{n+1} &= \vec{x}^* + \frac{1}{2}\vec{a} t_f^2,
	\end{align}
	for the acceleration, and meanwhile count the flux caused by the movement of particles, denoted as $\vec{w}_{i}^{fr,part}$,
	\begin{gather}
	\vec{w}_{i}^{fr,part} = \sum_{k\in P\left(\partial \Omega_{i}^{+}\right)} \vec{\phi}_k - \sum_{k\in P\left(\partial \Omega_{i}^{-}\right)} \vec{\phi}_k,
	\end{gather}
	where $P\left(\partial \Omega_{i}^{+}\right)$ is the particle set moving into the cell $i$ during one time step, $P\left(\partial \Omega_{i}^{-}\right)$ is the particle set moving out of the cell $i$ during one time step, $k$ is the particle index in the set, and $\vec{\phi}_k=\left[m_{k}, m_{k}\vec{u}_k, \frac{1}{2}m_{k}\vec{u}^2_k + m_k\frac{K+3}{2} \frac{1}{2\lambda_k}\right]^T$ is the mass, momentum and energy carried by particle $k$.
	\item Delete particles with $t_f < \Delta t$, and the carried conserved variables will merge into $\vec{W}$.
	
	Calculate the TKE characterized by the surviving particles, $\rho E_t$.
\end{enumerate}
Up to now, we have obtained all the necessary components of the flux, enabling the subsequent update of conserved quantities.
\begin{gather}\label{particle phase equ_updateW_ugkp}
\vec{W}_i^{n+1} = \vec{W}_i^n
- \frac{\Delta t}{\Omega_i} \sum_{S_{ij}\in \partial \Omega_i}\vec{F}^{eq}_{ij}S_{ij}
- \frac{\Delta t}{\Omega_i} \sum_{S_{ij}\in \partial \Omega_i}\vec{F}^{fr,wave}_{ij}S_{ij}
+ \frac{\vec{w}_{i}^{fr,part}}{\Omega_{i}}.
\end{gather}
It is worth noting that, the update of $\vec{W}_i^{p}$ inside each cell can be obtained by summing the contributions from all particles survived inside the cell, and further $\vec{W}^h$ can be obtained based on the conservation $\vec{W}^{h}_i = \vec{W}_i^{n+1} - \vec{W}^p_i$.
Besides, the unresolved TKE mainly evolves in WPTS based on the particles' evolution, such as sampling, movement, deletion, etc.

\begin{rmk}
In this paper, the statistically consistent calculation is employed, which means the $\vec{F}^{fr,wave}$ is reduced to $\vec{F}^{fr,UGKS}$ and systematically only the movement of surviving particles (from previous step) will contribute to $\vec{w}^{fr,part}$ while particles newly sampled at the begin of the current step will not be counted for $\vec{w}^{fr,part}$.
\end{rmk}

\begin{rmk}
From the particle's evolution, it can be understood that particles whose trajectories consistently pass through flow regions with high turbulence intensity exhibit a higher probability of survival at each step, and continuously transport flow information further downstream, consequently conveying flow features to far-downstream regions.
It directly demonstrates the non-local feature in turbulence modeling of the WPTS approach.
\end{rmk}

As discussed above, within the turbulence modeling framework of WPTS, particles move at their characteristic velocity, and the transport time should depend on the turbulence intensity along their trajectory. That is, in regions of high intensity, particles are more likely to remain active and keep moving downstream; conversely, in regions of low turbulence intensity, particles are more likely to vanish and return the carried conserved quantities back to the background wave component. Particularly, in WPTS the transport time is mainly determined by $\tau_t$. Therefore, it is necessary to develop the model of $\tau_t$, capable of reasonably reflecting the flow characteristics, such as the turbulence intensity, to accurately predict the turbulent flows.
In this paper, the model based on the mixing length hypothesis is developed for $\tau_{t} = \rho \nu_t / p$, with
\begin{gather}\label{nut}
\nu_t = \left(C_{ml}l\right)^2 |\vec{S}|,
\end{gather}
where $C_{ml} l$ is the modeled mixing length for jet flow
\begin{gather}\label{nut-len}
l = \sqrt{D x} + b_{ml},
\end{gather}
which increases with jet distance $x$. The term of velocity gradient in the original mixing length hypothesis is replaced by the strain rate tensor $\vec{S}$ in Eq.\eqref{nut}. Certainly, some previous investigations on the mixing length modeling have also utilized the strain rate or its simplified form rather than the original velocity gradient \cite{Tur-model-mixlen-shallow-stansby2003mixing}.

For the spatially developing jet flow, after the transition, the flow gradually evolves into a fully developed turbulent state. During this state, as the jet develops downstream, the overall turbulence intensity generally decays, accompanied by radial expansion. The $\tau_t$ from Eq.\eqref{nut} captures the overall decrease along the streamwise direction, with the coefficient $C_{ml}$ adjusting the decay rate. Besides, in the radial direction, due to the inherent flow features, the central region exhibits higher turbulence intensity and more small-scale structures, while the outer region approaches a laminar-like state with a smoother flow field. This typical flow structure inherently results in higher values of $\vec{S}$ in Eq.\eqref{nut} near the center and lower values toward the periphery, automatically satisfying the expected spatial distribution of $\tau_t$ for the jet flow.
Therefore, it can be expected that the WPTS coupled with the above $\tau_t$ model has the potential to provide reasonable predictions.

This model shares a form similar to the widely used Smagorinsky model (SM), with the modification being the replacement of the cell size $\Delta$ by the mixing length scale.
It is expected that this change will make the model more dependent on the typical feature of the flow pattern.
The jet flow has characteristic structures, where the overall morphology remains similar across Reynolds numbers.
Therefore, we expect the model given in Eq.\eqref{nut} holds the potential to achieve reasonably reliable predictive accuracy under different $Re$ with only minor adjustments to the model coefficients for example, which is highly attractive for practical applications and actually constitutes the primary motivation of this study. Thus, in the next section, the jet flows at two Reynolds numbers (5,000 and 20,000) are tested by WPTS equipped with the $\tau_t$ (or equivalently $\nu_t$) model to validate its performance.

\section{Numerical simulation}	
In this section, the numerical simulation for round jet flow by the WPTS method and results will be presented, including the computational setup, boundary conditions, the typical flow features, analysis, etc.

\subsection{Numerical and case setup}
As the treatments of WPTS in \cite{Tur-wpts-first-yang2025wave, Tur-wpts-second-yang2025wave}, the fifth-order WENO-AO reconstruction and second-order reconstruction with van Leer limiter are adopted for $\vec{W}$ and $\vec{W}^h$ respectively \cite{GKS-HLLC-compare-yang2022comparison, wenoao-gks-ji2019-performance-enhancement}.
For the temporal discretization, the two-step fourth-order method is employed for the wave component, while the stochastic particle component is transported once in a whole time step $\Delta t$.
Besides, it is worth noting that, in WPTS the flux $\vec{F}$ of GKS will be automatically employed \cite{GKS-2001} in the cell without stochastic particles, which ensures the exact fifth-order NS solution in this region.
For all simulations in this paper, the reference mass of one stochastic particle is $10^{-3}\Omega$. Besides, the CFL number is taken as 0.3, and $C_0$ in the modeled TKE production is taken as 0.5, which are the same as our previous studies of the turbulent jet flow
\cite{Tur-wpts-second-yang2025wave}.

As described in the introduction section, the main target of this paper is develop a model of $\tau_t$ for jet flow based on the mixing length hypothesis, and further test the performance of WPTS coupled with this new model. Now the definition of the jet Reynolds number $Re_j$ is given as
\begin{gather*}
Re_j = \rho U_e D / \mu_e,
\end{gather*}
where $\rho$ and $\mu_e$ are the density and viscosity of jet flow, respectively. Besides, $U_e$ is the core velocity of the jet, and $D$ stands for the diameter of the jet exit, taken as a unit. In all simulations in this paper, the viscosity is obtained by the power law, namely $\mu\left(T\right) = \mu_e \left(T/T_e\right)^{\omega}$ where $T_e$ is the temperature of jet flow and $\omega=0.667$.
Another dimensionless parameter is the Mach number ($Ma$), which is given by
\begin{gather*}
Ma = U_e / \sqrt{\gamma R T_e}.
\end{gather*}
In this paper, firstly the $Re_j$ with $5000$ and $Ma = 0.6$ is tested. Previous studies about the jet flow with $Re_j=5000$ through DNS can be found in  \cite{Tur-case-jet-DNS-sharan2021investigation}, which is taken as the DNS reference in this paper.

In the simulation, the whole computational domain has dimensions of $X\times Y\times Z$ of $45D \times 30D \times 30D$, which comprises a core simulation domain of $35D \times 20D \times 20D$ surrounded by an outer buffer zone.
The whole domain employs the non-uniform Cartesian grid with grid number $84^4$.
In detail, the smallest cell is located at the jet exit, and the cells are gradually expanding in the streamwise ($x$) and transverse ($y/z$) directions.
Specifically in the core domain, the minimum cell size in the streamwise direction $\Delta x_{min}$ is 0.15$D$, stretching with a fixed ratio of 1.025. Then in the transverse directions, the minimum cell size $\Delta y_{min} =\Delta z_{min}$ is 0.10$D$, stretching with a fixed ratio of 1.052.

The left boundary is set as the inflow boundary condition, where a velocity profile is given with superimposed perturbations.
Many previous studies based on DNS and LES have shown that the treatment of the inflow boundary is crucial for the accurate prediction of spatially developing jets. Otherwise, a systematic deviation, primarily characterized by the delay, in the distribution of streamwise velocity along the $x$-axis may occur.
To mitigate it, approaches are commonly adopted, such as adding inflow the perturbation of white noise or specific excitation modes \cite{Tur-case-jet-inflow-landa2004development, Tur-case-jet-inflow-gohil2015simulation, Tur-case-jet-inflow-gohil2019numerical},
using the pseudo turbulent inflow velocities \cite{Tur-case-jet-inflow-wang2010direct}, employing more realistic inflow data from the pipe flow \cite{Tur-case-jet-DNS-sharan2021investigation, Tur-case-jet-DNS-uPdfAna-nguyen2024analysis}, and modifying the subgrid stress model in LES \cite{Tur-case-jet-des-zhou2025enhanced}, etc.
In this work, we focus primarily on the self-similar characteristics of the jet flow, including mean flow variables and turbulent quantities such as Reynolds stresses. Detailed comparisons of the near-field flow, such as accurate prediction of the transition, will be addressed in future studies.
Therefore, the perturbation model developed in previous studies is employed, and the performance of WPTS coupled with the $\tau_t$ model developed based on the mixing length hypothesis, particularly the self-similar feature of fully developed turbulent jet flow, will be evaluated.

In this study, the mean velocity profile of the inflow is given by a hyperbolic tangent function:
\begin{gather*}
U(r)=\frac{U_e}{2}\left[1-\text{tanh}(\frac{r-r_0}{2\theta_0})\right], ~~ V=0, ~~W=0,
\end{gather*}
where $r_0 = D/2$ is the jet radius, $r = \sqrt{y^2+z^2}$ is radius, and $\theta_0$ is the initial momentum thickness taken as $0.04r_0$ \cite{Tur-case-jet-DNS-sharan2021investigation}.
Besides, for the excitation of turbulence, the dual-mode perturbations $u(r,t)$ developed in \cite{Tur-case-jet-inflow-gohil2015simulation} are superimposed on the streamwise velocity $U$ within the jet core $r\leq r_0$,
\begin{gather}\label{eq-uInletmode}
u(r,t)=A_n U_e \text{sin}\left(2 \pi St_D \frac{U_e}{D} t\right)
+ A_h U_e \text{sin}\left(2 \pi St_H \frac{U_e}{D} t - \theta \right)\left(\frac{2r}{D}\right),
\end{gather}
where $\theta = \text{atan}\left(y,z\right)$ is the azimuthal direction, the amplitudes are taken as $A_n=A_h=0.04$, the axisymmetric Strouhal number is $St_D = 0.5$, and the helical Strouhal number is $St_H = St_D / f$ with the frequency ratio $f = 2.40$. In addition, the small perturbations uniformly distributed in $[-kU_e, kU_e]$ are further superimposed on three velocity components in the jet core $r\leq r_0$ with $k = 0.004$ \cite{Tur-case-jet-inflow-gohil2015simulation}.
The convective and non-reflecting boundary conditions are imposed for the downstream and transverse boundaries, respectively.
Finally, it is worth noting that the simulation is conducted under the dimensionless system. Specifically, the length and velocity are non-dimensionalized by the jet diameter $L_{ref} = D$ and the velocity $U_{ref} = \sqrt{RT_e}$, respectively. Therefore, the dimensionless time is $t_{ref} = L_{ref} / U_{ref}$.

\subsection{Validation of WPTS with the mixing-length $\tau_t$ model at Re = 5,000}
\subsubsection{Typical flow features}
For the jet flow with Reynolds number 5000, $C_{ml}^2 = 6.1\times10^{-4}$ and $b_{ml} = 1.6$ in Eq.\eqref{nut} and Eq.\eqref{nut-len} are employed.
The numerical simulation spans a dimensionless time of $t=1660$, and the flow fields in $540\sim1660$ are used for the statistical analysis of turbulent quantities.

\begin{figure}[htbp]
	\centering
	\subfigure{
		\includegraphics[height=6.0cm]{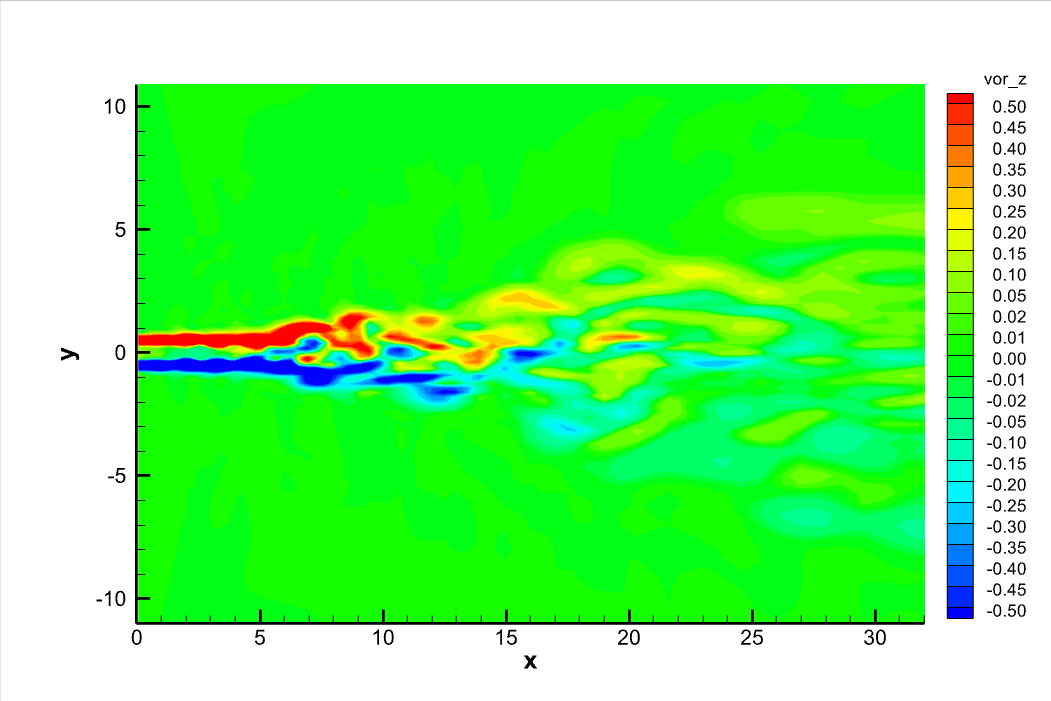}
	}
	\caption{The instantaneous snapshot of vorticity $\omega_z = V_x - U_y$ in WPTS at $t=1650$ for $xoy$ plane with $z=0$. The presented domain is around $[0,32D]\times[-11D,11D]$.}
	\label{Fig-vorz-z0}
\end{figure}

Firstly, the instantaneous flow field at $t = 1650$ is presented to illustrate the characteristics of the spatially developing jet. Figure \ref{Fig-vorz-z0} shows the distribution of the vorticity component $\omega_z$ on the $xoy$ plane at $z = 0$.
Although vorticity does not precisely reflect the distribution of turbulence intensity, it is often used as a key variable to characterize the overall flow features of turbulence. The results in Figure \ref{Fig-vorz-z0} indicate that after undergoing transition, the jet reaches a turbulence stage.
In this stage, the flow structures are gradually weakened as the flow develops downstream, accompanied by the radial flow expansion.
In regions far from the jet center (where $r$ is large), the flow becomes smooth, namely a laminar state. These typical flow characteristics are consistent with findings from previous numerical studies of jets using DNS.

\begin{figure}[htbp]
	\centering
	\subfigure{
		\includegraphics[height=5.6cm]{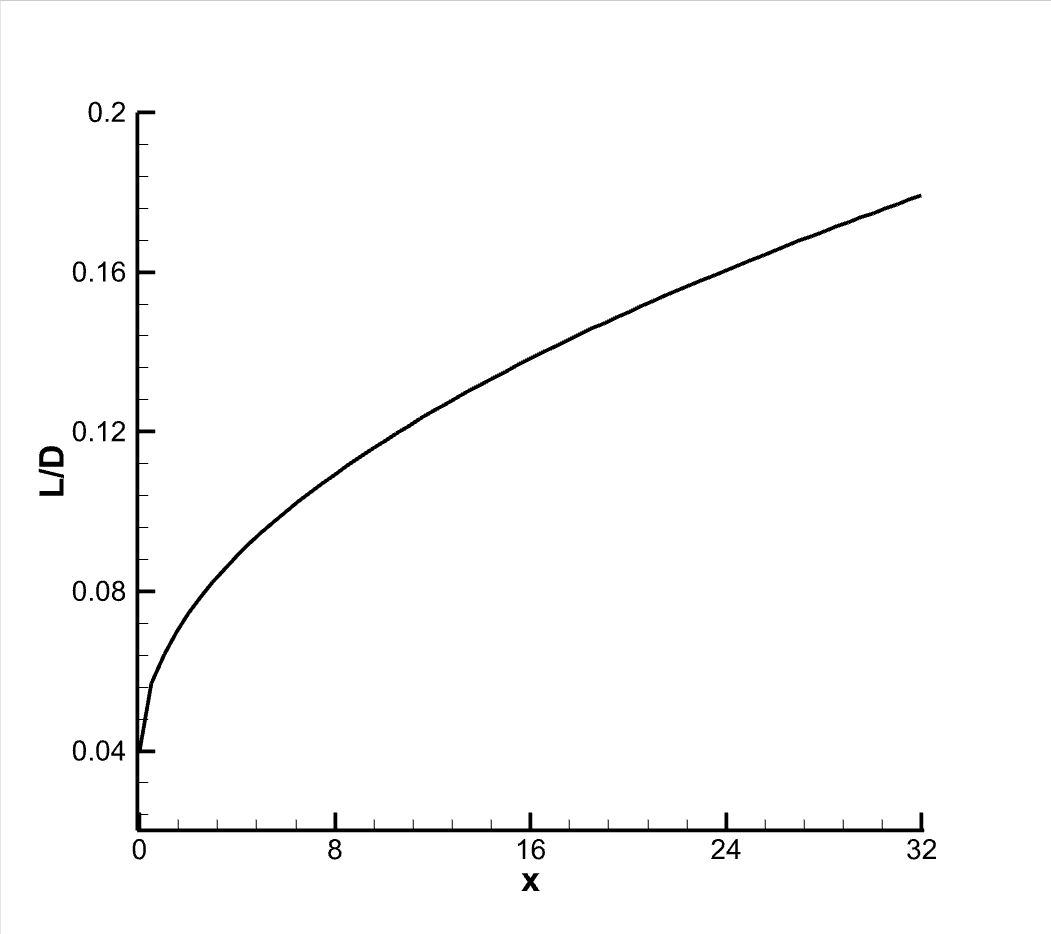}
	}
	\quad
	\subfigure{
		\includegraphics[height=5.6cm]{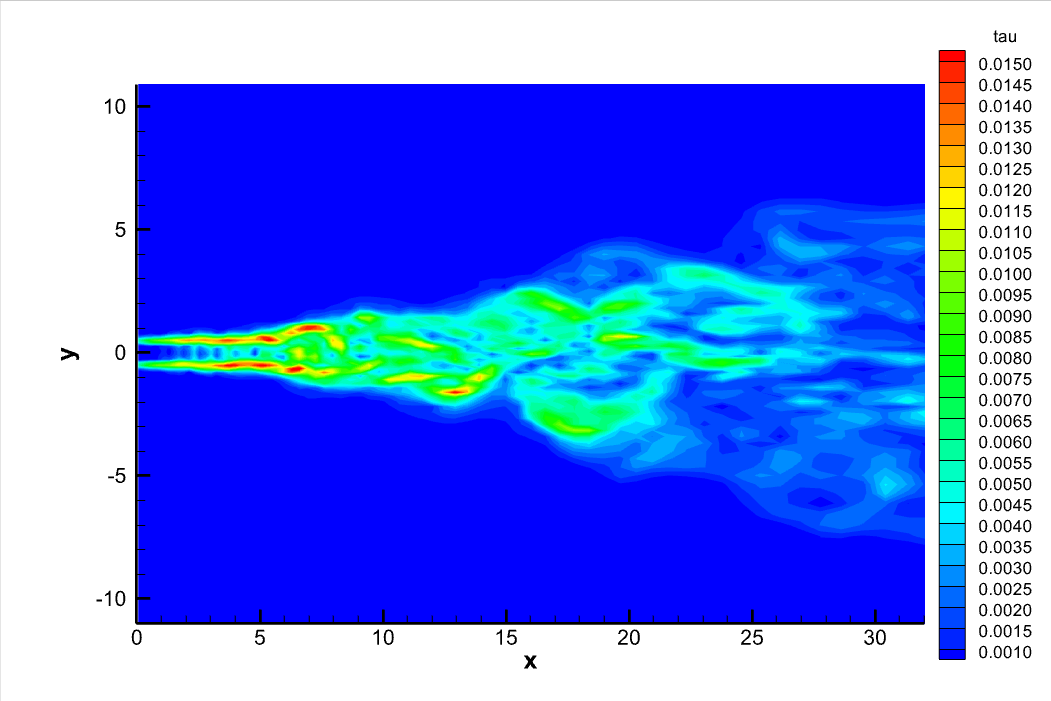}
	}
	\caption{(a) The modeled mixing length $C_{ml}l$ in Eq.\eqref{nut} with $C_{ml}^2 = 6.1\times10^{-4}$ and $b_{ml} = 1.60$, normalized by the jet diameter $D$. (b) The instantaneous snapshot of $\tau_n$, the sum of $\tau_p$ and $\tau_t$, at $t=1650$ for $xoy$ plane with $z=0$. The presented domain is around $[0,32D]\times[-11D,11D]$.}
	\label{Fig-tau-z0}
\end{figure}

As mentioned earlier, in WPTS the flow field is decomposed into wave and particle components, whose coupled evolution yields the predicted flow field. Specifically, the proportion of particles in WPTS is primarily governed by the turbulence characteristic time $\tau_t$.
As shown in Eq.\eqref{nut} and Eq.\eqref{nut-len}, in this study the $\tau_t$ model, based on the mixing-length hypothesis, depends mainly on $x$ and the cell-resolved strain-rate tensor $\vec{S}$.
Figure \ref{Fig-tau-z0}(a) shows the distribution of the modeled mixing length $C_{ml}l$, obtained based on Eq.\eqref{nut-len}, increasing with $x$.
Figure \ref{Fig-tau-z0}(b) presents the instantaneous distribution of $\tau_n$, the sum of the turbulence characteristic time $\tau_t$ and the physical value $\tau_p$ on the $xoy$ plane.
Since the modeled $\tau_t$ is typically significantly larger than $\tau_p$ in turbulent regions, the distribution of $\tau_t$ in Figure \ref{Fig-tau-z0}(b) essentially reflects that of $\tau_t$.
Overall, $\tau_t$ decreases with the increasing $x$ or $r$, which can be primarily attributed to the smaller magnitudes of the cell-resolved strain-rate tensor $\vec{S}$ in weakly turbulent and laminar regions, the flow feature that can be inferred from the vorticity field. In general, WPTS equipped with the $\tau_t$ model derived from Eq.\eqref{nut} and Eq.\eqref{nut-len} yields predicted jet flow characteristics that align with expectations, indicating the model's reasonableness.

\begin{figure}[htbp]
	\centering
	\subfigure{
		\includegraphics[height=7.0cm]{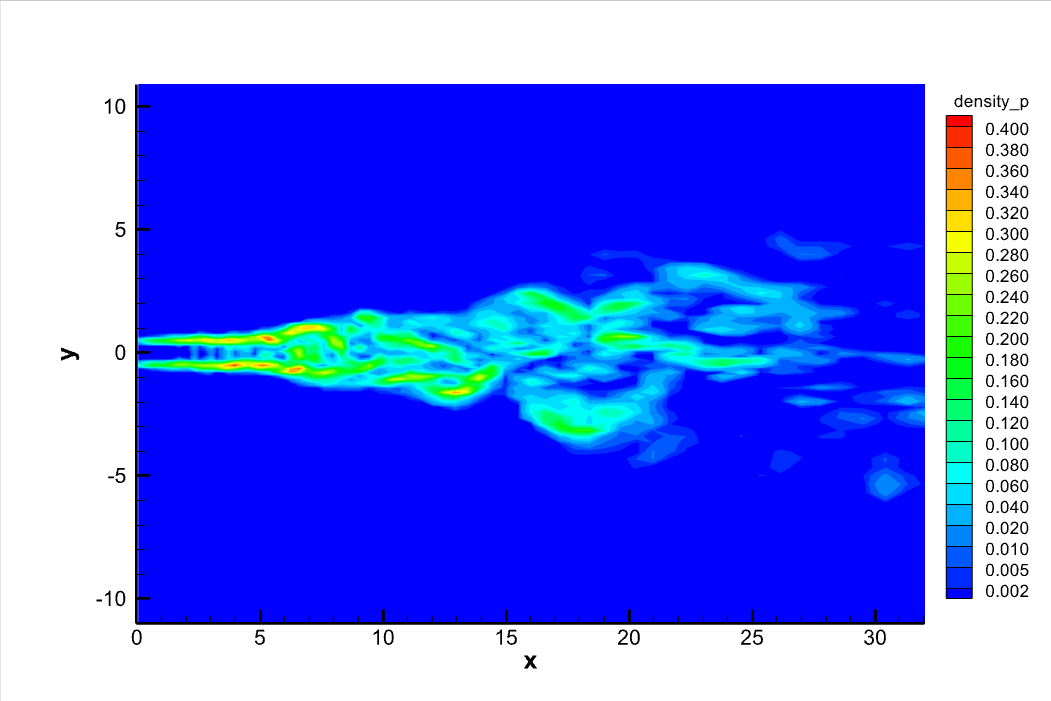}
	}
	\caption{The instantaneous snapshot of $\rho_p$, indicating the particle component in WPTS, at $t=1650$ for $xoy$ plane with $z=0$. The presented domain is around $[0,32D]\times[-11D,11D]$.}
	\label{Fig-rhop-z0}
\end{figure}

Furthermore, Figure \ref{Fig-rhop-z0} shows the instantaneous particle distribution in WPTS method at $t=1650$. It exhibits a high correlation with the distribution of $\tau_n$ shown in Figure \ref{Fig-tau-z0}(b), which can be attributed to the particle sampling strategy employed within WPTS.
Overall, particles mainly appear within the central flow with stronger turbulence of the jet. Their proportion gradually decreases as the flow develops downstream and expands radially.
Notably, in regions far from the jet center where the flow becomes laminar, particles automatically disappear, which ensures the computational efficiency, and more importantly, guarantees that the algorithm can fully recover to the NS solution in the laminar region.
It clearly demonstrates one of the advantages of the WPTS method, namely, the flow-regime adaptivity of wave-particle decomposition, achieving the unified description of both laminar and turbulent flows within the wave-particle framework.

Besides determining the particle proportion, $\tau_t$ in WPTS also serves to measure particle transport time, or equivalently migration distance.
As introduced, WPTS is a turbulence modeling and simulation method with typical non-local characteristics.
It is mainly because particles have the potential to evolve over multiple time steps or, equivalently, traverse multiple grid cells, thereby transmitting flow information far away.
In essence, the distribution of $\tau_t$ represents the local characteristic migration time given to particles during their transport process. Consequently, if a particle's trajectory remains in a region with a relatively large $\tau_t$, there is a high probability (but not absolute, as particles are reassessed for deletion at each time step) that the particle will survive multiple time steps and correspondingly migrate across multiple cells, carrying flow information downstream.

\subsubsection{Statistical variables}
\begin{figure}[htbp]
	\centering
	\subfigure{
		\includegraphics[height=6.5cm]{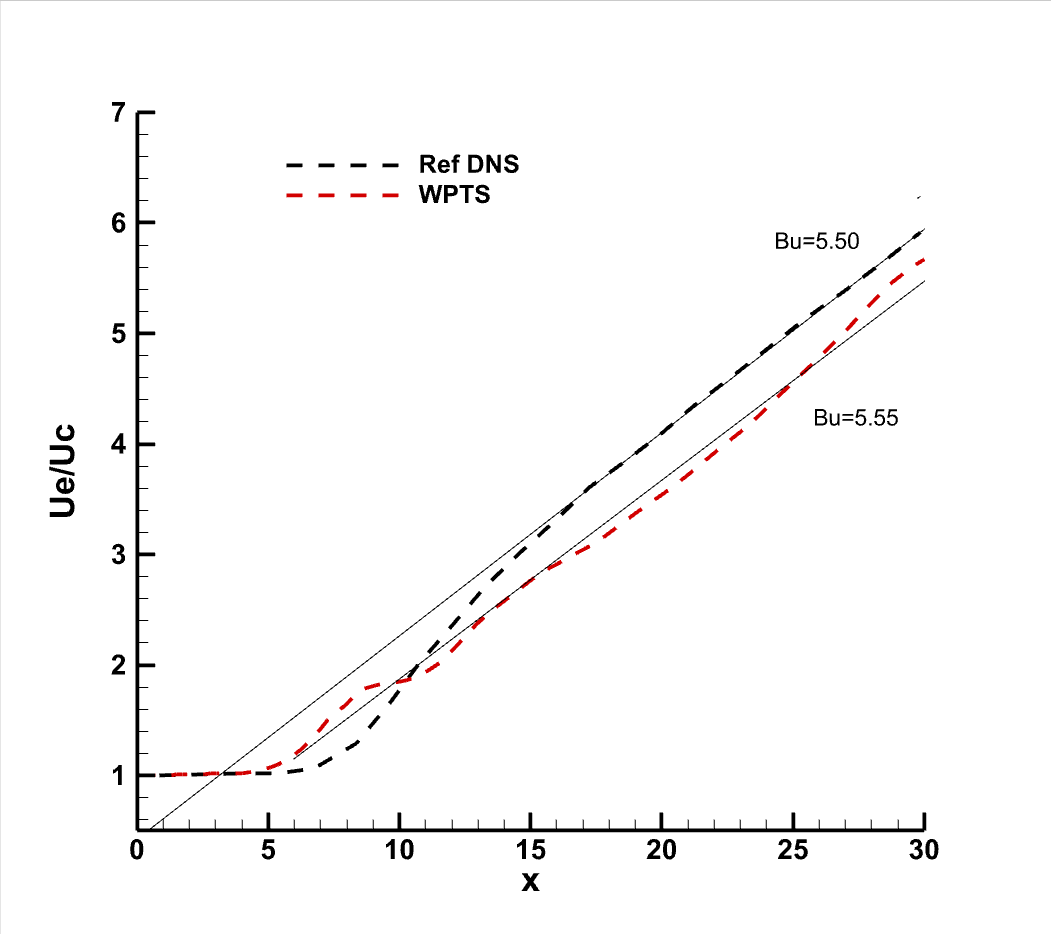}
	}
	\caption{The inverse of the averaged centerline mean velocity, and the reference is from \cite{Tur-case-jet-DNS-sharan2021investigation}.}
	\label{Fig-Bu}
\end{figure}

Firstly, the streamwise velocity $U_c(x)$ around the jet core ($r=0$) at different jet distances $x$ will be analyzed. Previous studies show that in the fully developed turbulence state of jet flow, the $U_c(x)$ can be evaluated as
\begin{equation}\label{Ucx}
\frac{U_e}{U_c(x)} = \frac{B_u D}{x - x_{0u}},
\end{equation}
where $U_e$ is the streamwise velocity at the jet core, $B_u$ and $x_{0u}$ indicate the decay rate of streamwise velocity and the assumed virtual origin of jet flow, respectively. It is worth noting that Eq.\eqref{Ucx} does not hold in the region near the jet exit (small $x$ region).
One typical feature of the fully developed turbulence is that the value $B_u$ shows a constant value at different $x$ positions, indicating the linear increase of $U_e/U_c(x)$.
More importantly, the $B_u$ are around one constant value independent of $Re$ \cite{pope2001turbulent}.
The predicted profile of $U_e/U_c(x)$ by WPTS is given in Figure \ref{Fig-Bu}, where the decay constant $B_u$ can be obtained, namely 5.55, which is a reasonable value compared with the reference value from experiment and DNS studies, given in Table \ref{Tab-Bu} \cite{Tur-case-jet-exp-hussein1994velocity, Tur-case-jet-exp-panchapakesan1993turbulence, Tur-case-jet-DNS-sharan2021investigation}.

\begin{table}[h]
	\centering
	\begin{tabular}{ccc}
		\hline
		& $Re_j$ & $B_u$ \\
		\hline
		Exp Hussein & 100,000 & 5.80 \\
		\hline
		Exp Panchapakesan & 11,000 & 6.06 \\
		\hline
		DNS Sharan & 5,000 & 5.50 \\
		\hline
		Present WPTS & 5,000 & 5.55 \\
		\hline
	\end{tabular}
	\caption{The predicted $B_u$ by WPTS with $\tau_t$ model based on mixing-length hypothesis.}
	\label{Tab-Bu}
\end{table}

\begin{figure}[htbp]
	\centering
	\subfigure{
		\includegraphics[height=6.0cm]{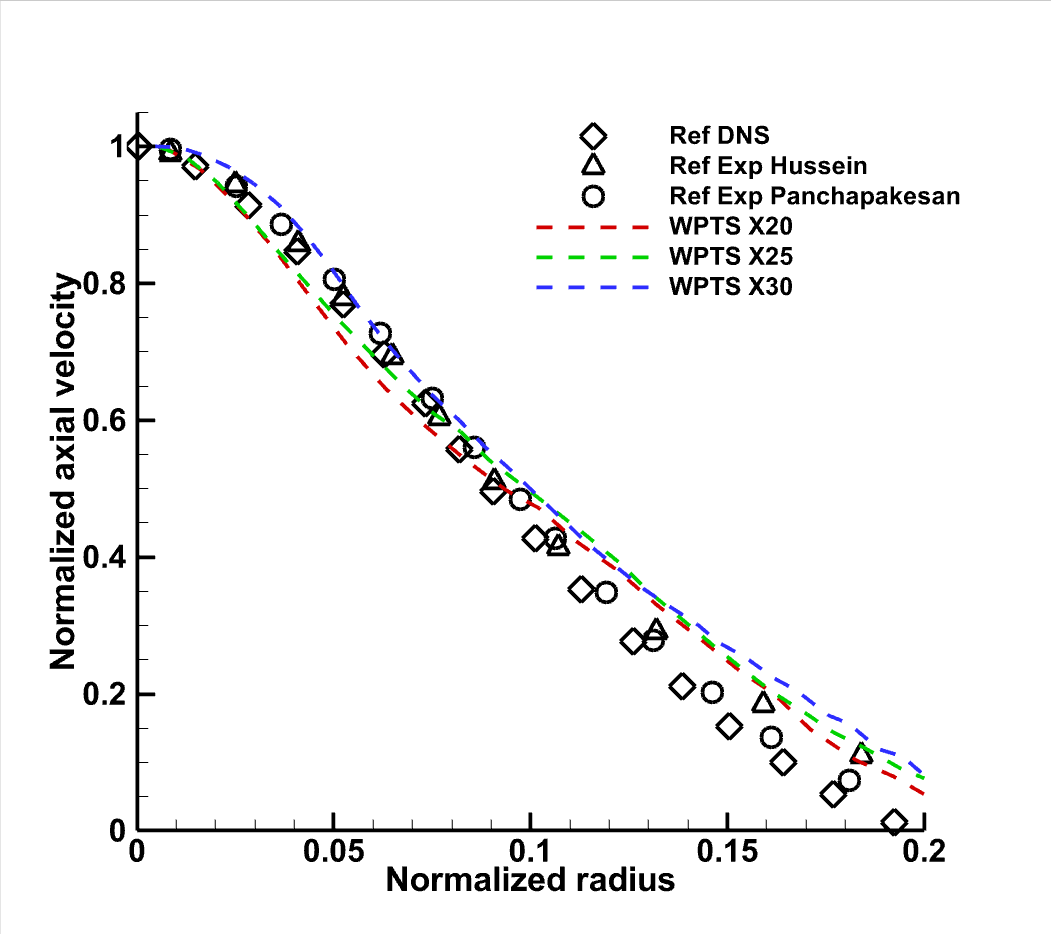}
	}
	\quad
	\subfigure{
		\includegraphics[height=6.0cm]{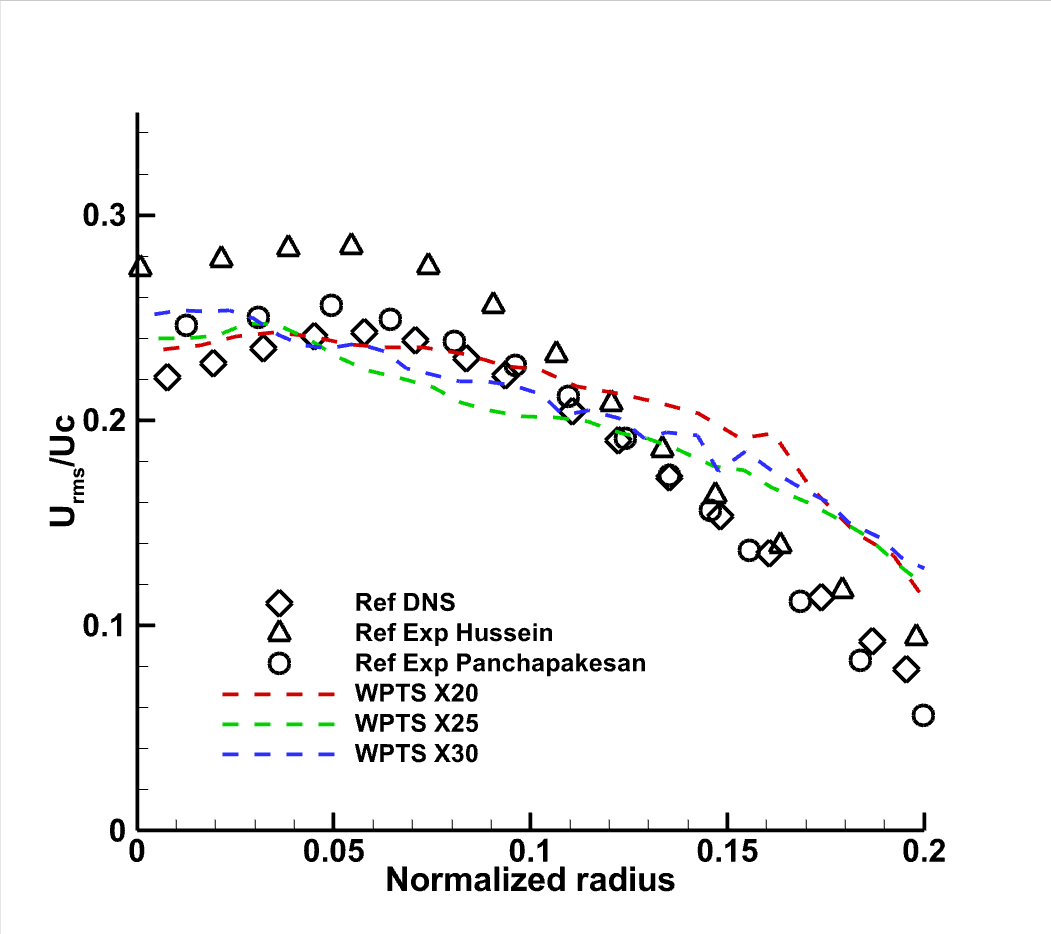}
	}
	\quad
	\subfigure{
		\includegraphics[height=6.0cm]{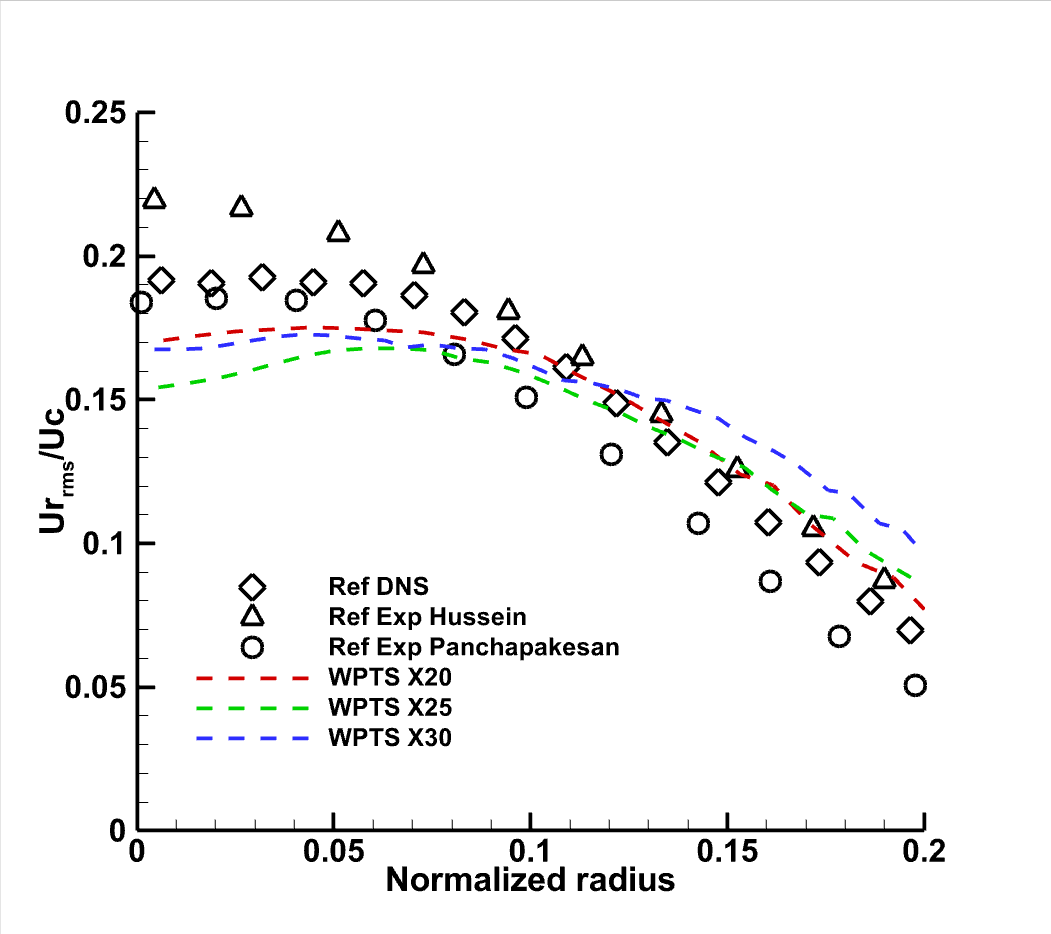}
	}
	\quad
	\subfigure{
		\includegraphics[height=6.0cm]{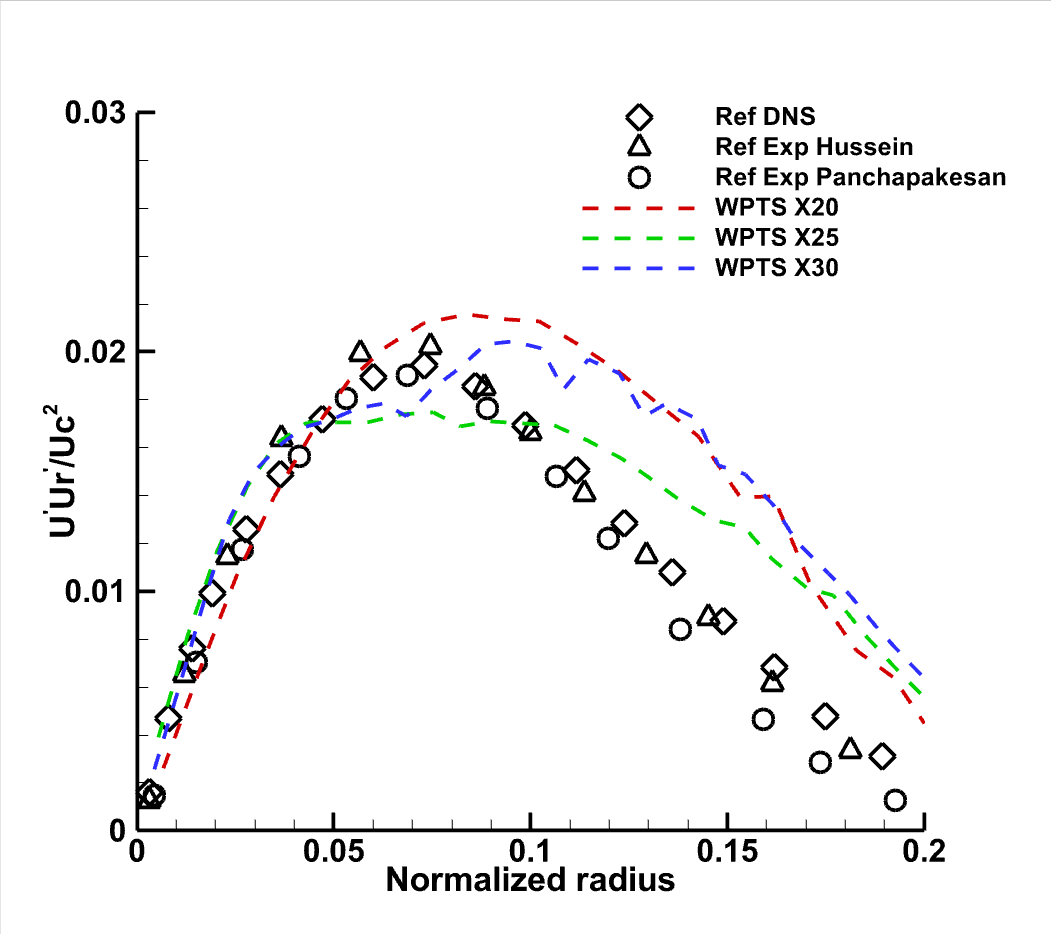}
	}
\caption{The profiles of (a) mean streamwise velocity, and Reynolds stress associated terms: (b) the r.m.s. of $U' U'$, (c) the r.m.s. of $U'_r U'_r$, (d) the cross-stress term of $U' U'_r$. $U$ and $U_r$ denote the velocity in the streamwise direction and radial direction, respectively.}
\label{Fig-vel-rxx}
\end{figure}

To obtain the turbulence statistic variables, the $V$ and $W$ velocity components in the Cartesian coordinate system need to be transformed into $U_r$ and $U_{\theta}$ velocity components in cylindrical coordinates.
Besides, the averaging in temporal and azimuthal directions is employed in getting the statistical variables along the radius, which are presented in Figure \ref{Fig-vel-rxx}.
Importantly, the statistical variables at different streamwise positions $x = 25, 30, 35$ across $10D$ are presented to validate their consistency, which is one significant feature in the self-similar flow pattern of fully developing jet turbulence.
The velocity and radius are normalized by the mean streamwise velocity at the centerline $U_c\left(x\right)$ and as $r_n = r/(x-x_{0u})$, respectively.
Further Figure \ref{Fig-vel-rxx}(b), (c), and (d) give the typical Reynolds stress-associated terms, namely the root mean square (r.m.s.) of $U'U'$, the r.m.s. of $U_r'U_r'$, and the cross-stress term $U'U_r'$, normalized by the $U_c^2(x)$.
It is worth noting that the profiles at different $x$ positions nearly collapse together close to the reference results, which is very challenging for turbulence modeling and simulations under coarse grids.

Finally, there exists some difference in the WPTS results presented in Figure \ref{Fig-vel-rxx} from the reference value in the zone $r_{nor}>1.2$. Specifically, WPTS predicts a kind of higher overall.
To the author's knowledge, this deviation may be caused by the modeled $\tau_t$ or the mixing length $C_{ml}l$, which play key roles in wave-particle decomposition and determine particles' transport time scale in WPTS.
As discussed above, the modeled mixing-length depends only on $x$ but not radius $r$, and thereby the radial regulation of $\tau_t$ relies only on the cell-resolved $\vec{S}$, which includes too little freedom in the modeling to exactly predict the flow fields in the radius direction.
Subsequent research could explore the further incorporation of radius-dependent modeling to enhance prediction accuracy. In summary, achieving such results above with only two free parameters remains notably commendable.

\subsection{Validation of Reynolds number 20000}
As introduced above, one of the objectives of this study is that the WPTS equipped with the newly developed $\tau_t$ model is capable of accurately predicting the typical flow characteristics of jet flow under different Reynolds numbers. Therefore, in this section, a higher Reynolds number of 20,000 is selected as a further test case, which presents significant challenges for the DNS approach.
Specifically, taking the identical computational framework to the $\mathrm{Re} = 5000$ case, but adjusting the coefficient in the $\tau_t$ model and the grid resolution employed, the WPTS method is expected to give satisfactory predictions for high-Reynolds-number jet turbulence.
Through systematic testing, the coefficient in $\tau_t$ model is taken as $C^2_{ml}=3.5\times10^{-4}$, and also the mesh grid is refined to $\Delta x_{min}=0.114D$, while all other settings are kept the same with those in Reynolds number $5000$ case, including the $Ma$, boundary conditions, employed CFL number, coefficients in WPTS, etc.
It should be further noted that, once the minimum cell size $\Delta x_{min}$ is determined, the entire grid in the core domain will be uniquely determined, with the generation law that all these grids (including $\Delta x$, $\Delta y$, and $\Delta z$) are scaled by the refining factor determined through $\Delta x_{min}$. The grid number employed for the Reynolds number 20000 case is $100^3$.

\begin{figure}[htbp]
	\centering
	\subfigure{
		\includegraphics[height=5.2cm]{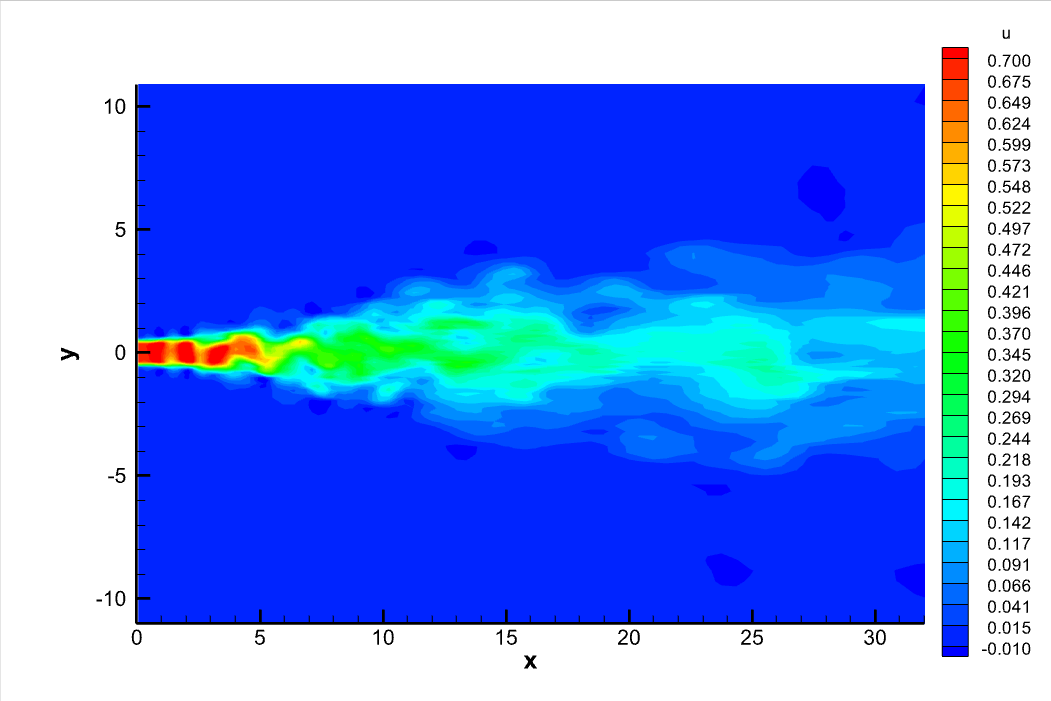}
	}
	\subfigure{
		\includegraphics[height=5.2cm]{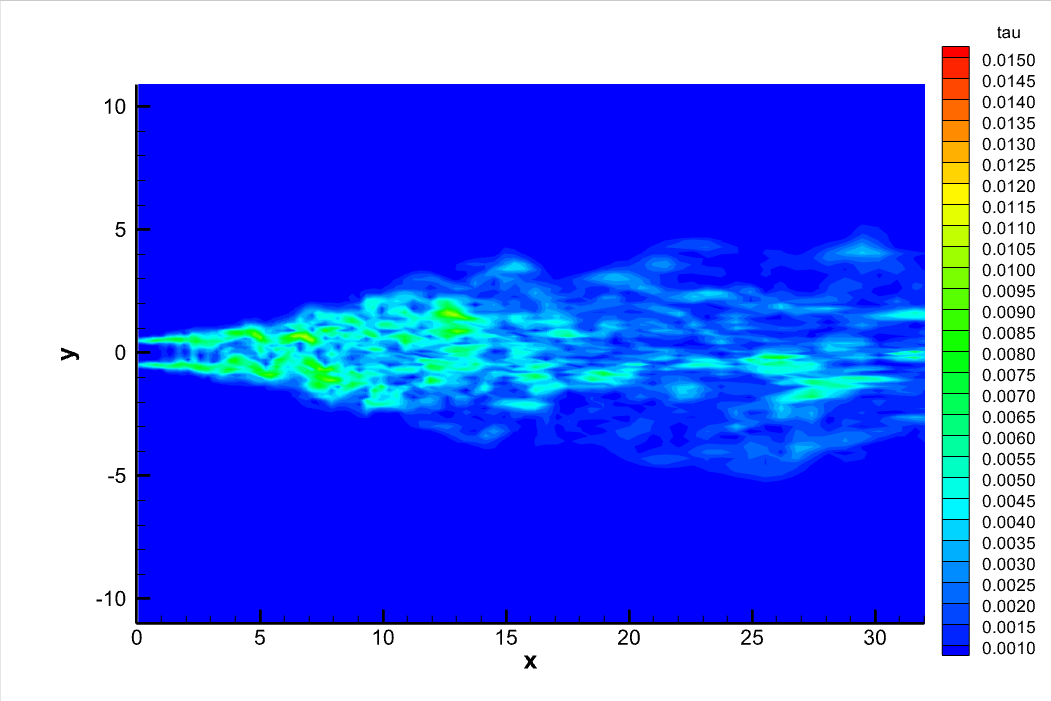}
	}
	\subfigure{
		\includegraphics[height=5.2cm]{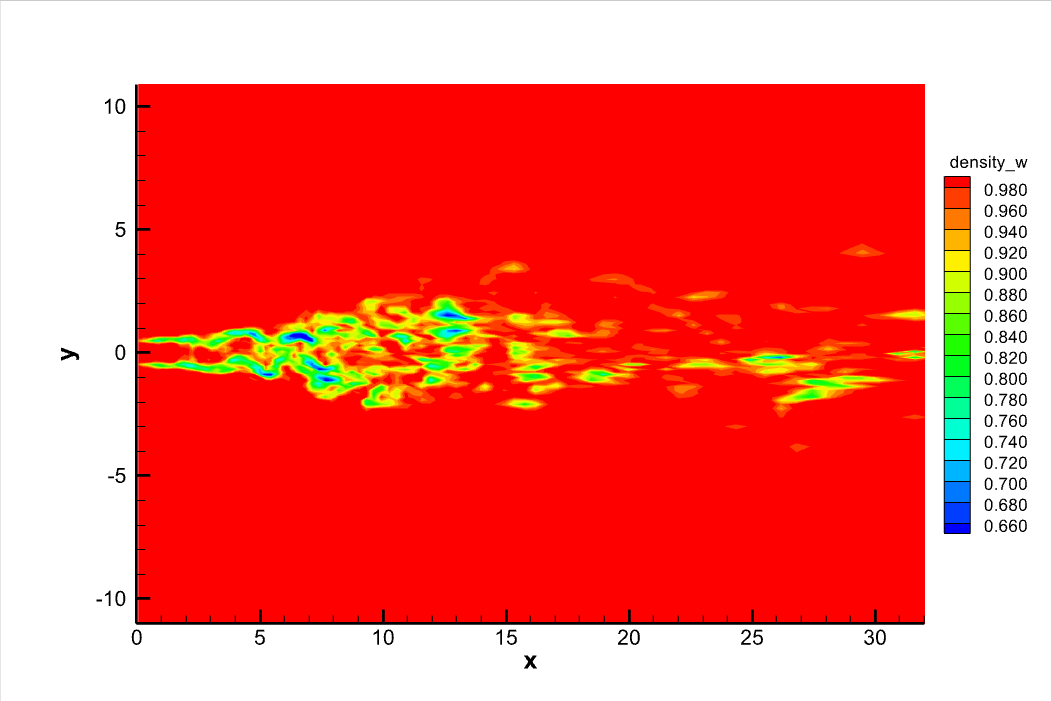}
	}
	\subfigure{
		\includegraphics[height=5.2cm]{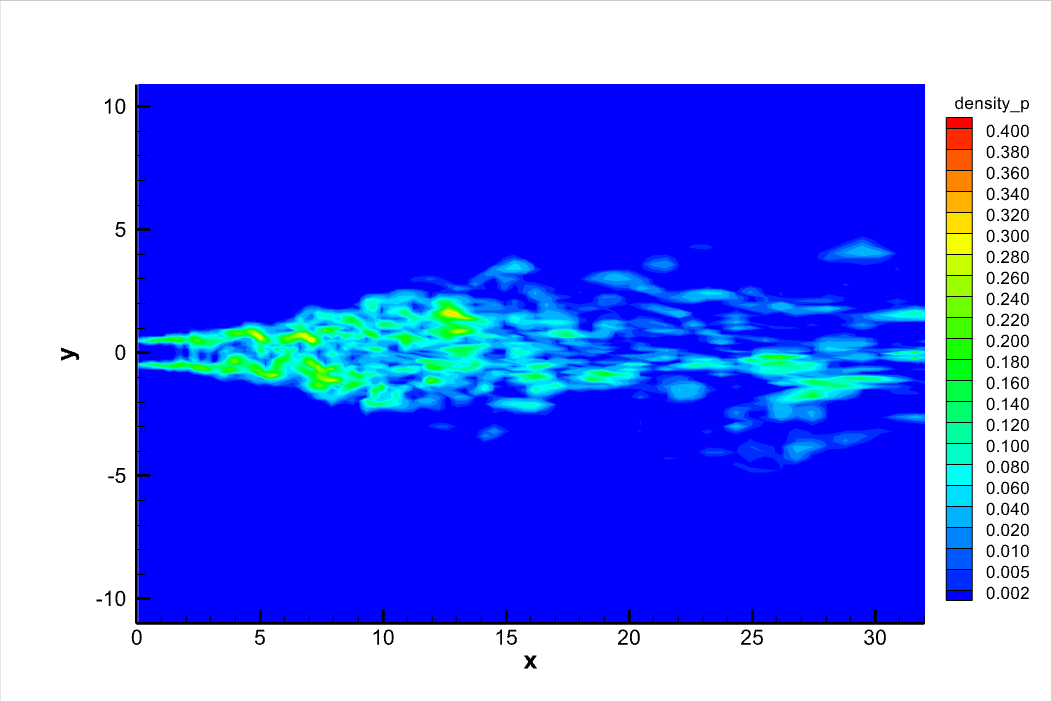}
	}
	\caption{The predicted flow fields of $Re=20000$ by WPTS. The instantaneous snapshots of (a) streamwise velocity, (b) total characteristic time, $\tau_p$ + $\tau_t$, (c) density of wave component in WPTS, and (d) density of particle component in WPTS.}
	\label{Fig-ins-20k}
\end{figure}

Firstly, Figure \ref{Fig-ins-20k} presents the instantaneous flow fields by WPTS of the Reynolds number $20000$ case, such as the streamwise velocity in Figure \ref{Fig-ins-20k}(a), and the total characteristic time $\tau_n = \tau_p + \tau_t$ in Figure \ref{Fig-ins-20k}(b).
The overall flow features of the jet flow are clearly observed: the flow expands radially as it develops downstream; the turbulence intensity shows a higher distribution in the jet center than the periphery zone, and gradually decays in the streamwise direction.
Further, Figure \ref{Fig-ins-20k}(c) and Figure \ref{Fig-ins-20k}(d) present the flow density of wave component and particle component in WPTS, namely $\rho_h$ and $\rho_p$, respectively. Generally, the higher value of turbulence collision time directly determines the longer transport distance of fluid elements, potentially producing the non-local feature in turbulence, which is achieved by sampling and tracking the stochastic particles in WPTS. While for the flow with low turbulence intensity and even in far laminar regions, the percentage of stochastic particle components decreases, and thus the wave component will dominate in the evolution.

\begin{figure}[htbp]
	\centering
	\subfigure{
		\includegraphics[height=6.0cm]{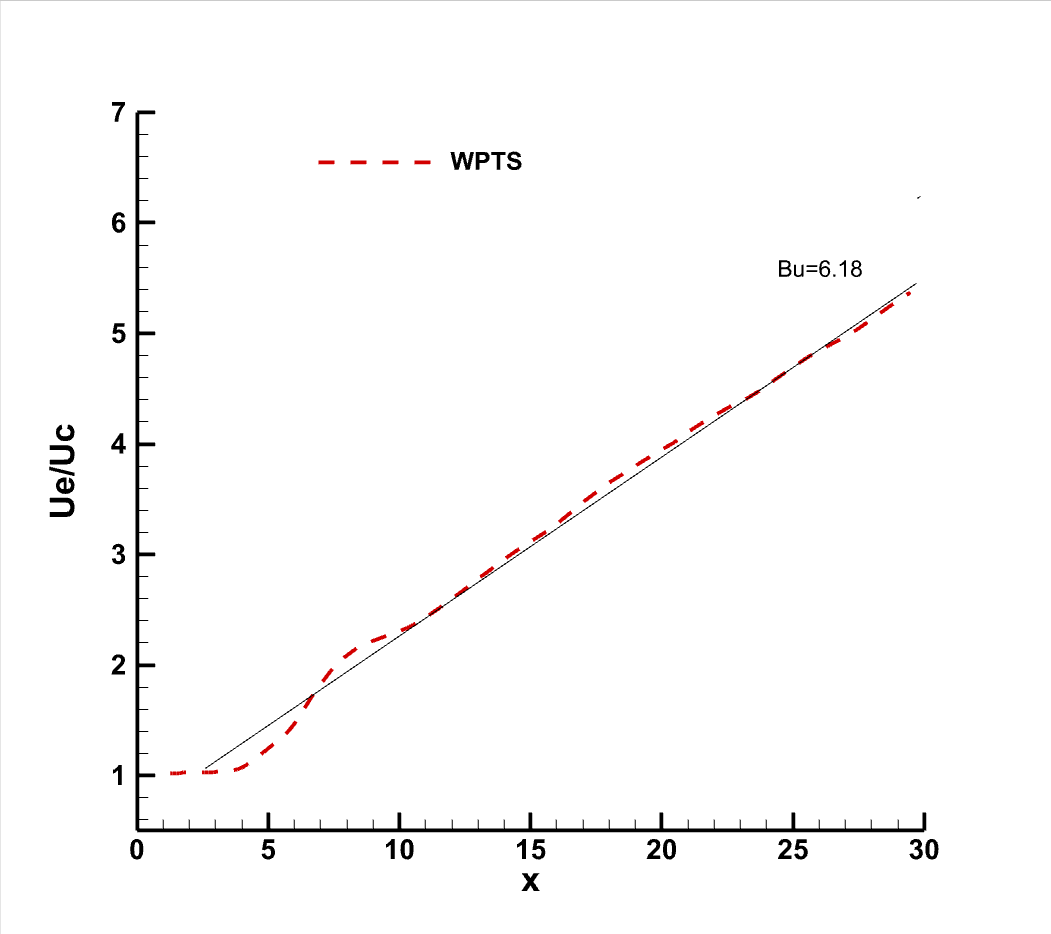}
	}
	\quad
	\subfigure{
		\includegraphics[height=6.0cm]{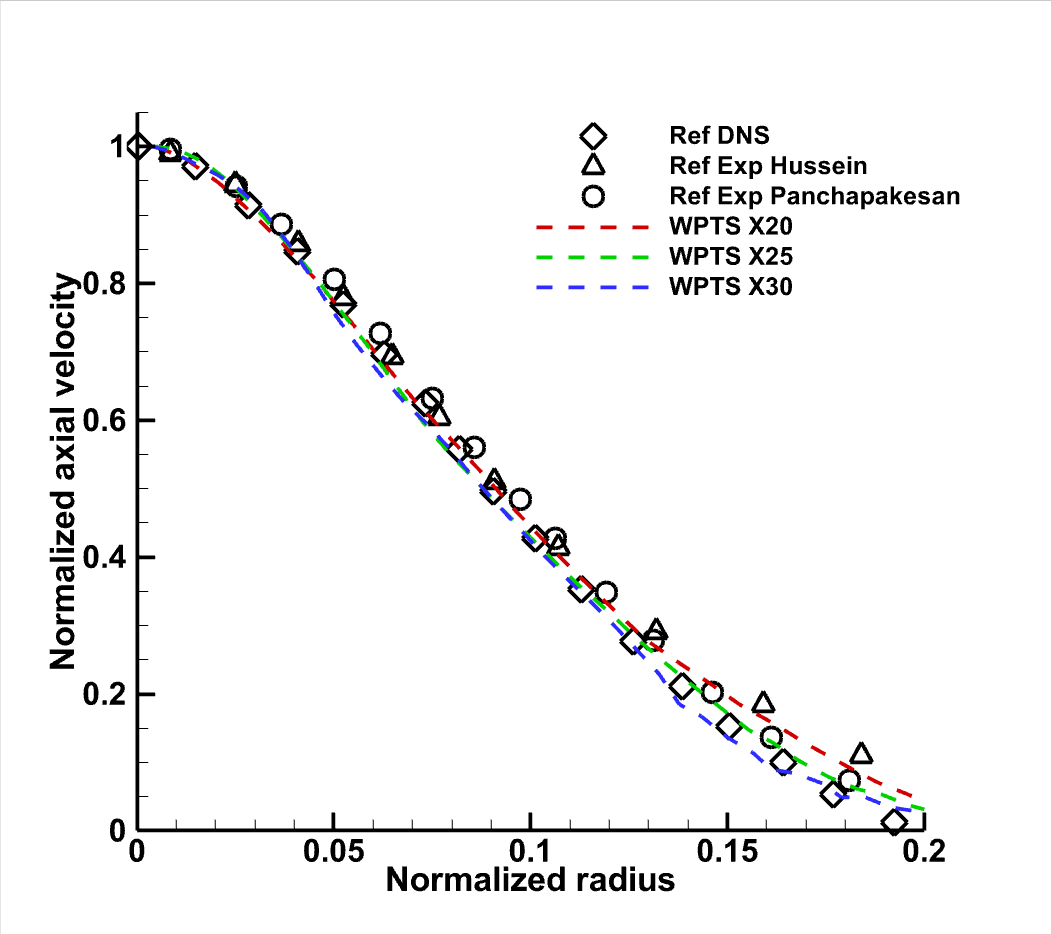}
	}
	\quad
	\subfigure{
		\includegraphics[height=6.0cm]{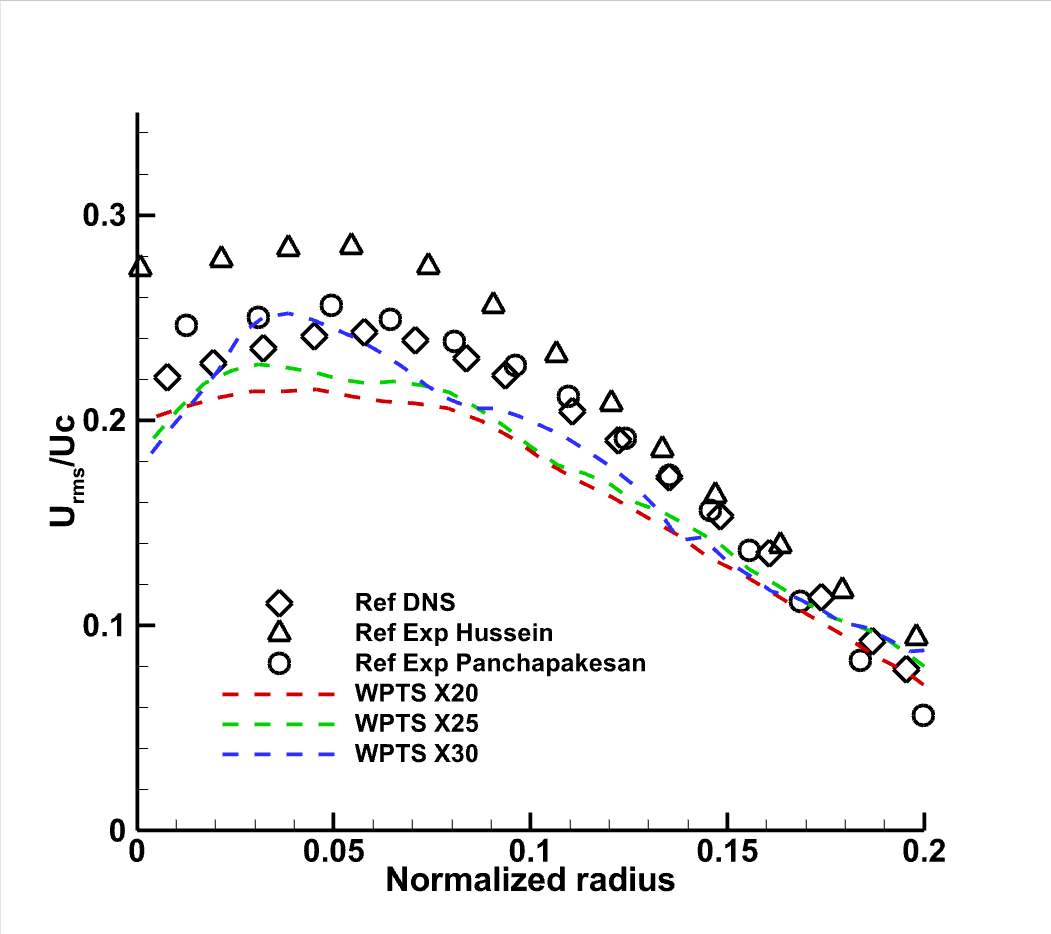}
	}
	\quad
	\subfigure{
		\includegraphics[height=6.0cm]{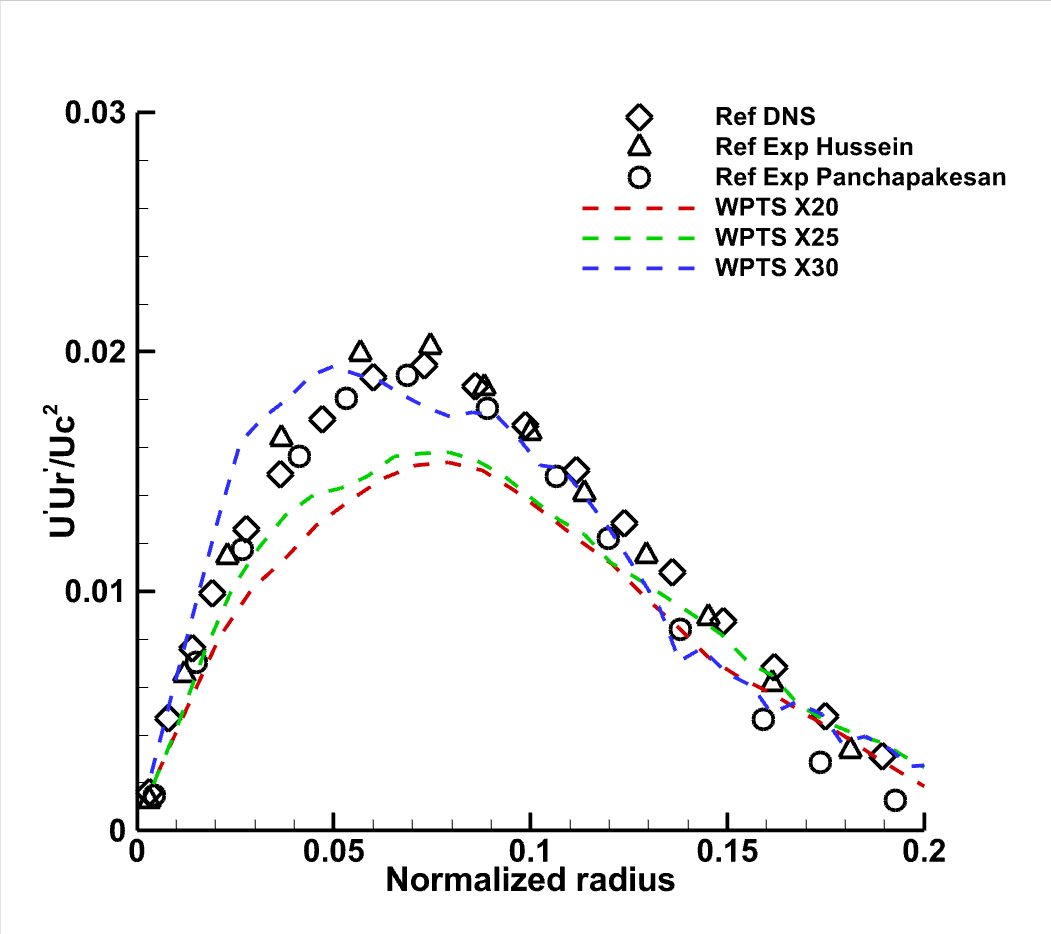}
	}
	\caption{The results of $Re=20000$ by WPTS. The profiles of (a) the $Ue/Uc$, (b) mean streamwise velocity, and Reynolds stress-associated terms: (c) the r.m.s. of $U' U'$, (d) the cross-stress term of $U' U'_r$. $U$ and $U_r$ denote the velocity in the streamwise direction and radial direction, respectively.}
	\label{Fig-velrxx-20k}
\end{figure}

Furthermore, the quantitative comparisons are presented in Figure \ref{Fig-velrxx-20k}, which effectively reflects the similarity flow feature of the fully developed jet turbulence.
Particularly, Figure \ref{Fig-velrxx-20k}(a) demonstrates that the decay feature of the mean streamwise velocity along the jet distance, presenting a distinct linear trend, with the slope value of $6.18$, deviating about $2\%$ from the reference value 6.06 in Panchapakesan's experimental measurement for the Reynolds number 11000 case \cite{Tur-case-jet-exp-panchapakesan1993turbulence}.
Additionally, Figure \ref{Fig-velrxx-20k}(b-d) presents the predicted distributions of the mean velocity and Reynolds stress components along the radial direction by WPTS equipped with $\tau_t$ constructed based on the mixing length hypothesis. Remarkably, quantities at different axial positions, namely $x=20D, 25D, 30D$, collapse well onto one single curve again for the Reynolds number 20000 case. This observation provides evidence for the accuracy and reliability of the method developed in this paper in predicting jet turbulence under different Reynolds numbers, from 5000 to 20000 for instance.
Finally, by comparing the case 20000 to 5000, the Reynolds-number-dependent scaling strategy can be summarized as: the grid size needs reduce to $Re_0^{-0.20}$, and the coefficient $C_{ml}^2$ in the $\tau_t$ model is scaled by $Re_0^{-0.40}$, where $Re_0$ is the ratio of Reynolds number to 5000, leading to 4.0 for case $Re=20000$. This Reynolds-number-dependent approach demonstrates the potential of the proposed method for the extension to other Reynolds number cases.

\section{Conclusion}

In this paper, we develop a turbulence characteristic-time model based on Prandtl’s mixing-length hypothesis and couple it with our recently proposed wave–particle turbulent simulation (WPTS) method to simulate spatially developing round jets. The resulting $\tau_t$ closure depends primarily on the streamwise distance
$x$ and the local strain-rate tensor $\vec{S}$. The WPTS formulation inherits the central physical picture of the mixing-length hypothesis: in regions of intense turbulence, discrete fluid elements travel a finite distance at a characteristic velocity before interacting with the surrounding fluid. In WPTS, this concept is realized by sampling stochastic particles in strongly turbulent regions and evolving them in a coupled manner with the background flow represented by the wave (Eulerian) component.

Despite this conceptual connection, the present approach differs fundamentally from eddy-viscosity models derived from mixing-length theory in two respects. First, particle free transport in WPTS explicitly retains non-equilibrium transport effects. Second, through coupled wave–particle evolution, WPTS provides a unified framework capable of representing both laminar and turbulent regimes. We first test the proposed method for a spatially developing jet at Re=$5,000$. With appropriate model coefficients, the method accurately reproduces the self-similar behavior of the fully turbulent jet, including the streamwise decay of the centerline velocity, mean velocity profiles at multiple downstream locations, and key turbulence statistics such as Reynolds-stress distributions. To examine robustness across Reynolds numbers, we further increase the jet Reynolds number to Re=$20,000$. The results indicate that WPTS continues to predict the jet self-similarity characteristics accurately with only two adjustments: (1) scaling the model coefficients according to a $-0.4$ power law of the Reynolds-number ratio, and (2) refining the minimum grid scale according to a $-0.2$ power law of the same ratio. Here, the Reynolds-number ratio is defined relative to the baseline Re=$5,000$ case, i.e., $4.0$ for Re=$20,000$.

\section{Acknowledgements}
The current research is supported by National Key R\&D Program of China (Grant Nos. 2022YFA1004500), National Science Foundation of China (12172316, 92371107), and Hong Kong research grant council (16301222, 16208324).

\bibliographystyle{plain}%
\bibliography{reference}
\end{document}